\documentclass[aps,pra,twocolumn,showpacs,preprintnumbers,amsmath,amssymb, superscriptaddress]{revtex4}

\usepackage{graphicx}
\usepackage{dcolumn}
\usepackage{bm}
\usepackage{color, float}

\usepackage{subfigure}

\newcommand{\pname}{${\rm ED}^3$}

\usepackage{multirow}

\usepackage{amssymb}
\usepackage{amsmath}

\begin{document}

\preprint{xxx}

\title
{
Anomaly detection in reconstructed quantum states using a
machine-learning technique
}

\author{Satoshi Hara
}
\thanks{These two authors contributed equally}
\affiliation{The Institute of Scientific and Industrial Research, Osaka University,
Mihogaoka 8-1, Ibaraki, Osaka 567--0047, Japan
}%
\affiliation{%
IBM Research - Tokyo, 5-6-52 Toyosu, Koto-ku, Tokyo 135-8511, Japan
}

\author{Takafumi Ono
}
\thanks{These two authors contributed equally}
\affiliation{%
Research Institute for Electronic Science, Hokkaido University,
Sapporo 060--0812, Japan
}%
\affiliation{%
The Institute of Scientific and Industrial Research, Osaka University,
Mihogaoka 8-1, Ibaraki, Osaka 567--0047, Japan
}%

\author{Ryo Okamoto}
\affiliation{%
Research Institute for Electronic Science, Hokkaido University,
Sapporo 060--0812, Japan
}%
\affiliation{%
The Institute of Scientific and Industrial Research, Osaka University,
Mihogaoka 8-1, Ibaraki, Osaka 567--0047, Japan
}%

\author{Takashi Washio}
\email{washio@ar.sanken.osaka-u.ac.jp}
\affiliation{%
The Institute of Scientific and Industrial Research, Osaka University,
Mihogaoka 8-1, Ibaraki, Osaka 567--0047, Japan
}%

\author{Shigeki Takeuchi}
\email{takeuchi@es.hokudai.ac.jp}
\affiliation{%
Research Institute for Electronic Science, Hokkaido University,
Sapporo 060--0812, Japan
}%
\affiliation{%
The Institute of Scientific and Industrial Research, Osaka University,
Mihogaoka 8-1, Ibaraki, Osaka 567--0047, Japan
}%

\date{\today}
%

\begin{abstract}
\noindent
The accurate detection of small deviations in given density matrices is important for quantum information processing. Here we propose a new method based on the concept of data mining. We demonstrate that the proposed method can more accurately detect small erroneous deviations in reconstructed density matrices, which contain intrinsic fluctuations due to the limited number of samples, than a naive method of checking the trace distance from the average of the given density matrices. This method has the potential to be a key tool in broad areas of physics where the detection of small deviations of quantum states reconstructed using a limited number of samples are essential.
\end{abstract}

\maketitle
%
\section{Introduction}

The field of quantum information processing is growing very rapidly. 
In addition to quantum computation and quantum key distribution\cite{Niels}, 
new applications like quantum simulations\cite{Guzik2012} and quantum metrology\cite{Nag2007,Oka2008} are attracting considerable attention. In quantum 
information processing, the detection of deviations from the normal state, or `errors', 
is a crucial task, as in classical information processing. Such `error' 
detection is also very important when detecting a target state which 
is close to the background (normal) states.
However, such error detection is very difficult because of the intrinsic statistical 
nature of quantum physics. Usually the state of a system, or the density matrix, 
is estimated from a limited number of experiments/samples using quantum 
state tomography\cite{Jam2001}. Therefore, the matrix elements of the density 
matrices have intrinsic fluctuations, making it difficult to distinguish 
erroneous states from normal states.

In this paper, we propose to apply `data mining' method to solve this
problem. Data mining is a computational process of discovering patterns 
in large data sets involving methods of machine learning, statistics, and
other methods. 
We focus on the detection on anomalies where the amplitudes of the 
elements of density matrices are different. Such anomalies include 
dephasing in the quantum states, which is very important in many 
applications and tasks of quantum information processing. 
For this purpose,
we have developed a new data mining method named 
`\pname \ (Erroneous Deviation Detection for Density matrices)' for 
quantum density matrices by extending a previously reported method\cite{hara2013}. 
We compare \pname \ and a naive method, in which the erroneous 
states are distinguished by the trace distance from the average of the 
states. We show that \pname \ has a significant advantage over the naive 
method for numerical simulation data and real experimental data. 
Our method can be applied to any quantum states represented by not only photonic qubits but also other many physical systems including superconducting circuits, trapped ions, and so on.
We believe quantum state data mining will be a key tool in the broad area of quantum computation, quantum communication, quantum metrology, 
and also in broad area of sciences where the detection of erroneous 
quantum states is critical.

%
\section{Methodology}
We consider $K$ state density matrices $\hat{\rho}_k \in \mathbb{C}^{d \times d}$ with 
$k = 1, \ldots, K$, each of which is obtained by quantum state tomography for a limited number of samples. 
Because now we are interested in the anomaly detection of amplitudes of 
the elements in the density matrices, which includes the dephasing of 
the quantum states, it is not necessary to detect the phases of the off-diagonal matrix elements.
Thus
we consider the absolute of the density matrix; 
each $(i, j)$th element of the matrix is given by $|\hat{\rho}_{k, ij}|$ where 
$\hat{\rho}_{k, ij}$ denotes the $(i, j)$th element of $\hat{\rho}_k$. We use the notation $\hat{\rho}_k$ to express the absolute matrix for brevity. 
Our task is to find $M$ erroneous matrices that are different from the remaining 
$K - M$ normal matrices. Here, we assume that the number $M$ is unknown 
but smaller than $K/2$, {\it i.e.}, erroneous observations are rare in 
experiments. This task is known as `anomaly detection' or 
`outlier detection' in data mining\cite{hodge2004}. 

\subsection{Naive Approach}

A natural way to solve this problem is to use a physical model for 
computing the state density matrix, which represents the normal experimental 
conditions, and find erroneous matrices that do not match the computed 
normal matrix. This requires a `true' physical model with `true' parameters 
representing the normal experiments. However, such true models and 
parameters are rarely known in most cases, even if  we 
presume an `ideal' experiment, because true experiments deviate 
from ideal experiments by some tolerated errors in their settings and some 
disturbance from their surrounding environment. Thus, this approach is not 
easily applicable in practice. 

An alternative naive method is to adopt the `data average'  
$\bar{\rho} \equiv \frac{1}{K} \sum_{k=1}^K \hat{\rho}_k$ as an approximation 
of the normal density matrix and represent each $\hat{\rho}_k$  
by a sum of $\bar{\rho}$ and its deviation $\tilde{\omega}_k$, {\it i.e.},  
$\hat{\rho}_k = \bar{\rho} + \tilde{\omega}_k$. 
We then measure the discrepancy of each density matrix $\hat{\rho}_k$ 
from $\bar{\rho}$ by $\tilde{\omega}_k =\hat{ \rho}_k - \bar{\rho}$. We apply the 
{\it trace distance}\cite{Niels2} as a convenient measure of the 
discrepancy as 
\begin{align}
	e_k \equiv \| \tilde{\omega_k} \|_{\rm tr} ,
	\label{eq:td}
\end{align}
where $\| * \|_{\rm tr}$ denotes the trace norm of a matrix, or the sum of the 
singular values. If $e_k$ is large, the target matrix is deemed to be erroneous. 
A drawback of this naive method is a reduced detectability of the errors 
$e_k $, because $\bar{\rho}$ is biased from the normal density matrix by 
the average of the matrices including the erroneous ones.

\subsection{Proposed Method}

In contrast to the naive approach, here we propose a method that considers an `ideal realization' of the normal 
matrix $\theta$ instead of $\bar{\rho}$ and an `ideal deviation' $\omega_k$ 
instead of $\tilde{\omega}_k$. That is, we adopt $\theta + \omega_k$ as an 
ideal density matrix for the $k$th experiment. If the $k$th experiment is 
conducted under normal conditions, the ideal realization is $\theta$, and 
the deviation $\omega_k$ becomes zero. On the other hand, a change in 
experimental conditions potentially creates a density matrix different from 
$\theta$ where the deviation $\omega_k$ may no longer be zero, 
indicating that the data is erroneous. The task is therefore to identify which 
density matrix $\hat{\rho}_k$ has a non-zero deviation $\omega_k$. Once 
such an $\omega_k$ is derived, we can measure the degree of error using 
Eq.(\ref{eq:td}). Formally, we expect the following two points to apply for the 
proper estimator of $\theta$ and $\omega_k$: 1) the observed matrix 
$\hat{\rho}_k$ is sufficiently close to the true realization $\theta + \omega_k$; 
2) the deviation parameter $\omega_k$ becomes zero under normal 
experimental conditions.

To reflect these two points in a sophisticated manner, we introduce a 
technique proposed in graphical Gaussian modeling (GGM), which is a 
research field of data mining.  The technique involves efficiently decomposing 
a set of precision matrices, which are inverse covariance matrices, into their 
common invariant elements and individually deviated elements\cite{zhang2010, hara2013}. 
Because these techniques are dedicated to the precision matrices, which 
are typically sparse and positive semi-definite (PSD), it is not applicable 
to our absolute density matrices, which are often dense and not limited to being
PSD. Accordingly, we newly formalize a task to decompose $\hat{\rho}_k$ 
into $\theta$ and $\omega_k$ over $k = 1, 2, \ldots, K$ as a `regularized' 
least squares regression problem:
\begin{align}
	\min_{\theta, \{\omega_k\}_{k=1}^K} \sum_{k=1}^K \frac{1}{2} \| \hat{\rho}_k - (\theta + \omega_k) \|_{\rm F}^2 
		+ \gamma \sum_{k=1}^K \sqrt{\sum_{i, j=1}^d s_{ij}^2 \omega_{k, ij}^2} \, ,
	\label{eq:qtad}
\end{align}
where $\| * \|_{\rm F}$ denotes the Frobenius norm of a matrix and $\gamma$ 
is a non-negative parameter used for optimization. 
Note that Frobenious norm is just used as a kind of mathematical tool to simplify the optimization problem, since this formulation makes it a simple convex problems.
The weight parameters 
$s_{ij}$ are given by the following for each element:
\begin{align*}
	s_{ij}^2 \equiv \left( \frac{1}{K} \sum_{k=1}^K \tilde{\omega}_{k, ij}^2 \right)^{-1} .
\end{align*}
The first term of Eq.(\ref{eq:qtad}) reflects the first point, that an observed 
matrix $\hat{\rho}_k$ has to be close to $\theta + \omega_k$. The second 
term, which reflects the second point, is the so-called `regularization' term. Intuitively, 
this term penalizes too large values of $\omega_k$ and reduces the estimator 
to be sufficiently small. Moreover, this term has the effect of reducing some of 
the deviations $\omega_k$ to be exactly zero for sufficiently large $\gamma$\cite{yuan2006}. 
The weight parameters $\{s_{ij}\}_{i,j=1}^d$ are introduced to balance the 
difference in scaling across the matrix entries. The effect of entries with 
large variations is suppressed by small $s_{ij}$, while the effect of entries with 
subtle deviations is magnified by the weighting. Note that this formalization 
does not require the matrices to be PSD.

\begin{figure}[b]
\begin{center}
\includegraphics*[width=8.5cm]{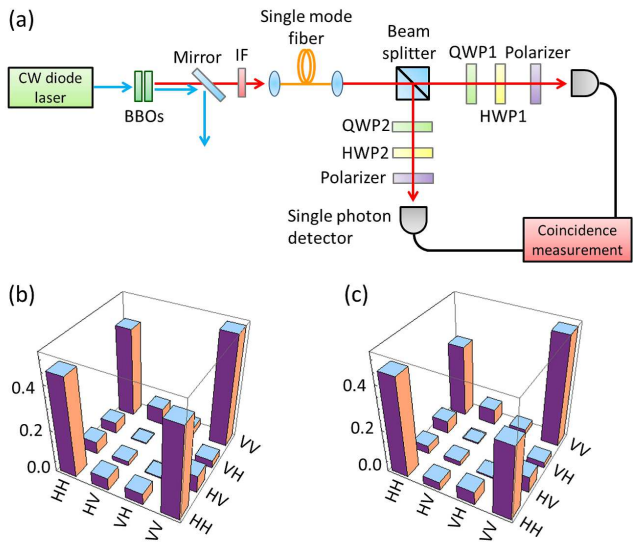}
\caption{(a) Schematic of the experimental setup for tomography.
An interference filter (IF) with 4-nm bandwidth was used. 
The tomography was implemented using quarter wave plates (QWP), half wave plates (HWP) and polarizers. 
(b) Histogram of the average of 300 density matrices for the normal states. 
(c) Histogram of the average of 50 density matrices for the erroneous states.
}
\end{center}
\label{fig:hist_simu_org}
\end{figure}

\begin{figure*}[t]
\centering
\subfigure[Raw elements and distances by the naive method]{
\includegraphics*[width=6.9cm]{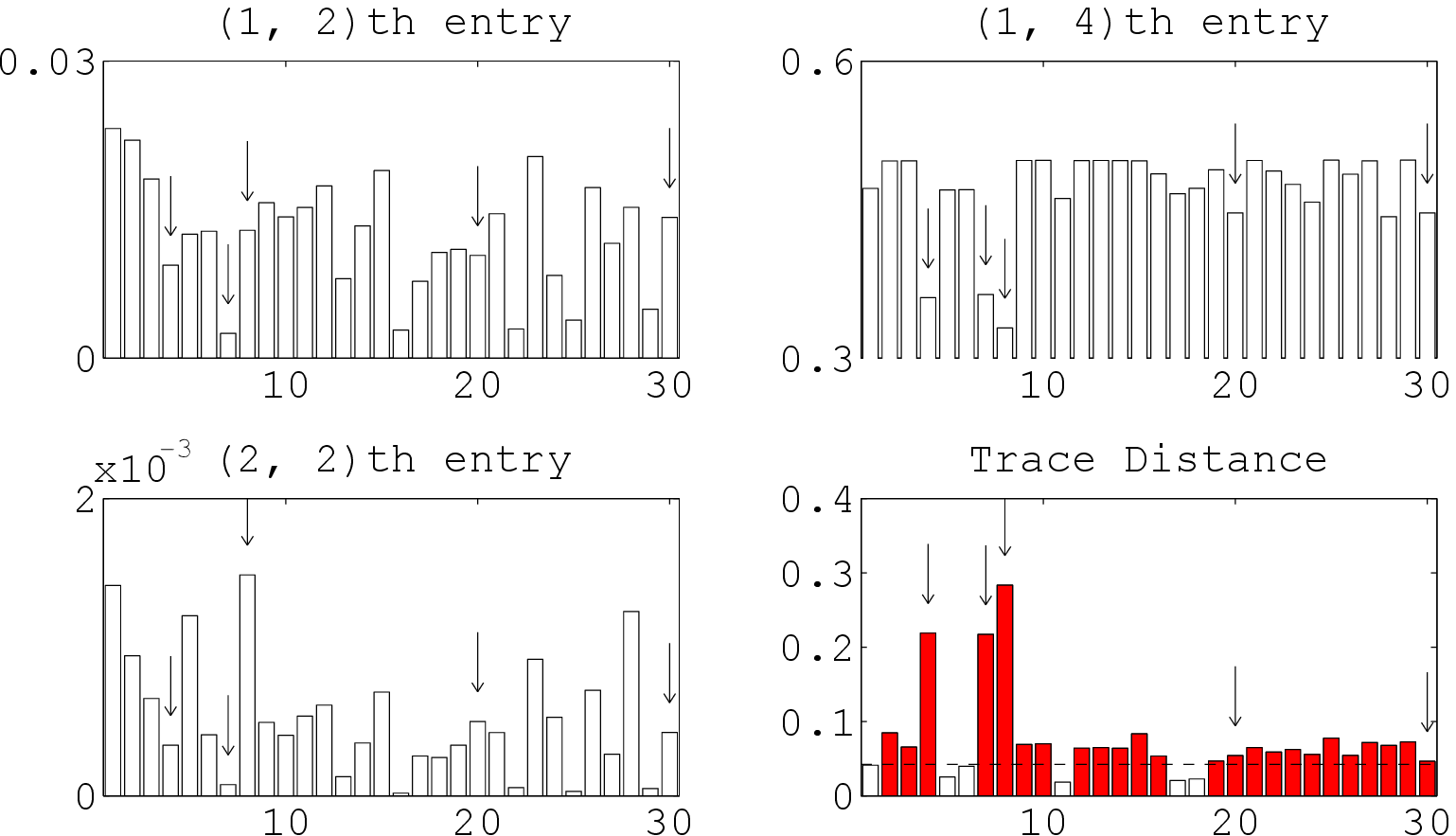}
\label{fig:hist_simu_org}}
\subfigure[Estimated elements and distances by \pname]{
\includegraphics*[width=6.9cm]{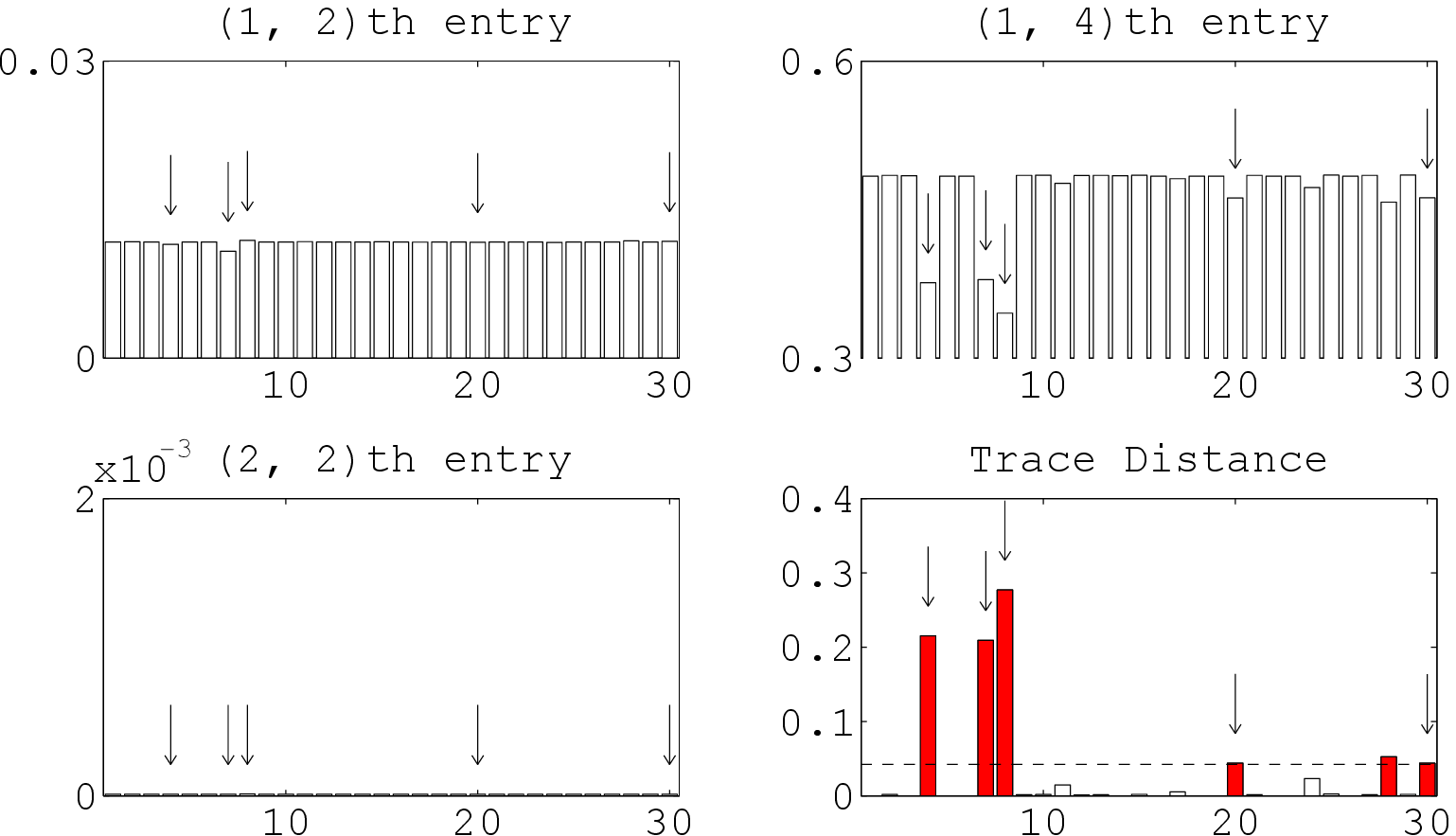}
\label{fig:hist_simu_proposed}}
\subfigure[ROC Curve]{
\includegraphics*[width=3.2cm]{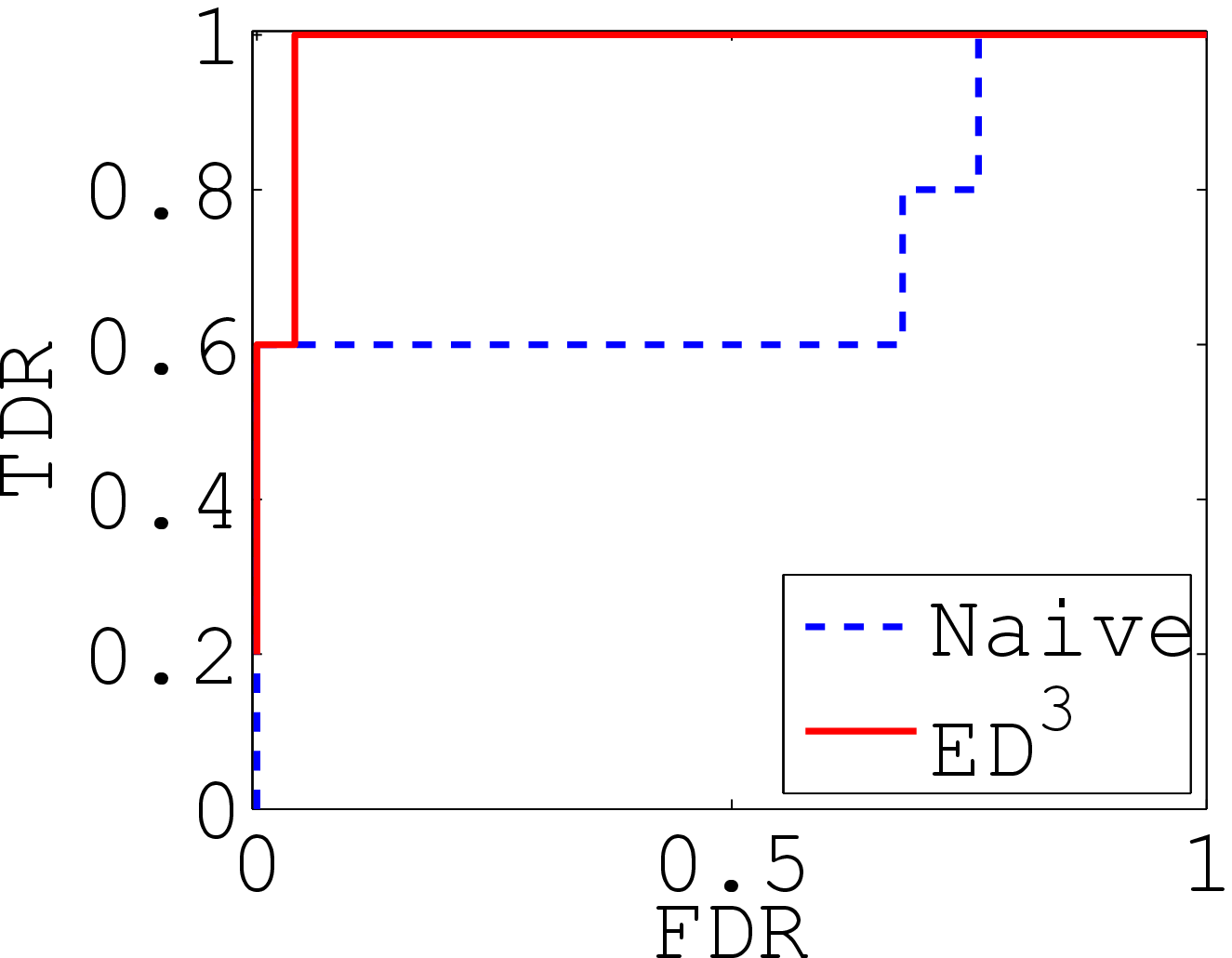}
\label{fig:roc_simu}}
\caption{Performance analysis using a computer-generated dataset. (a) The raw 
values of the elements (1,2), (1,4) and (2,2) and  and the trace 
distance of the 30 density matrices in a test dataset. The arrows denote 
the artificially
introduced erroneous cases. (b) Corresponding elements and the trace 
distance of the estimated density matirices using proposed method (\pname).
The colored trace distances represent the cases where the state density 
matrices are judged to be erroneous under a threshold level.
(c) The ROC curves for the naive method and \pname.}
\end{figure*}
\begin{figure}
\begin{center}
\includegraphics*[width=8.5cm]{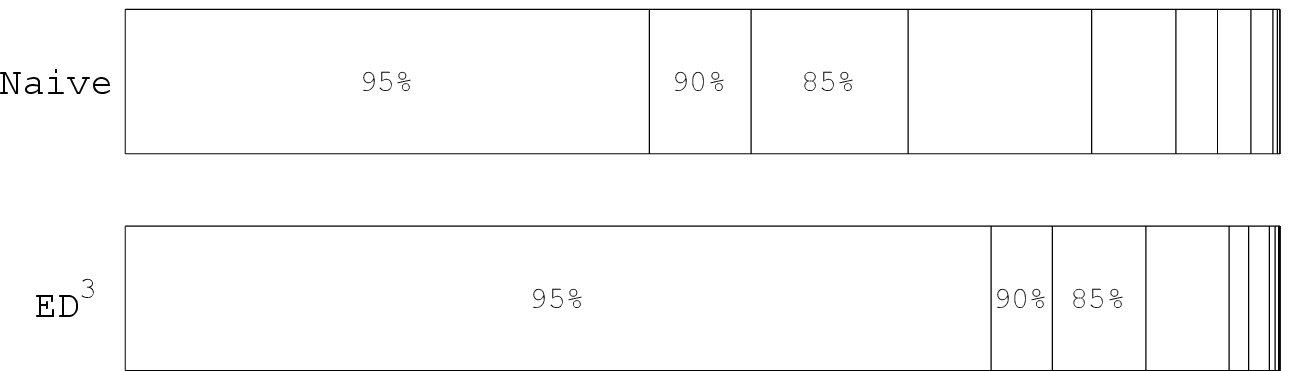}
\caption{Area under curve distribution over 1000 computer-generated datasets.}
\end{center}
\label{fig:bar_auc_simu}
\end{figure}

We also note that the formulation (\ref{eq:qtad}) is one specific example of the `sparse additive model'\cite{ravikumar2007spam} developed in machine learning.
In particular, this can be casted as a modification of the model presented in \cite{jalali2010dirty} where we replaced the regularization term that fits to our objective.
The advantage of these formulations is its high utilities;
they have several favorable theoretical properties as well as a computational tractability.
That is, the problem (2) is a convex optimization problem and the global minimum exists \cite{yuan2006}, 
and the parameter $\omega_k$ estimated by solving the problem is guaranteed to be sparse for sufficiently large $\gamma$ \cite{yuan2006}.
Hence, thanks to the convexity, we can use efficient methodologies developed in the optimization studies to solve the problem \cite{boyd2011}.
The computational tractability is particularly important for our application where the dimensionality of the density matrix grows exponentially as the number of photon increases.
Here, we describe an overview of the algorithm, which works up to several qubits  or for matrices of size around hundreds. Similar to our previous study\cite{hara2013}, we do not 
work on the problem (\ref{eq:qtad}) directly but work on the dual problem instead:
\begin{align*}
	& \min_{\{\zeta_k\}_{k=1}^K} \frac{1}{K} \sum_{k=1}^K \| \hat{\rho}_k - s \circ \zeta_k \|_{\rm F}^2 , \\
	& \text{subject to} \; \sum_{k=1}^K \zeta_k = 0_{d \times d} , \\
	& \hspace{48pt} \| \zeta_k \|_{\rm F}^2 \leq \gamma^2 \; (k = 1, 2, \ldots, K) ,
\end{align*}
where $s$ is a matrix that has $s_{ij}$ for the $(i, j)$th element and $\circ$ 
denotes a Hadamard product of matrices. From the duality, the optimal 
dual parameter $\zeta_k^*$ relates to the optimal primal parameters 
$\theta^*$ and $\omega_k^*$ as $\theta^* + \omega_k^* =\hat{\rho}_k - s \circ \zeta_k^*$.
Therefore, solving the problem (\ref{eq:qtad}) amounts to finding 
the optimal dual parameters $\{\zeta_k^*\}_{k=1}^K$. This can be conducted 
by properly reformulating the dual problem and applying the method called Alternating Direction Method of Multipliers (ADMM) method\cite{boyd2011}. 
This is our proposed approach, which we call `\pname \ (Erroneous Deviation Detection 
for Density matrices)'.

%
\section{Experimental Setup}

In the following, we compare the performance of \pname \ and the aforementioned 
naive method through experiments. We try to discriminate the density matrices of 
the `normal state', for which we use a two-photon polarization entangled state, from those 
of the `erroneous state', which slightly decoheres from the normal state. Note that the 
elements of the normal and erroneous density matrices have intrinsic fluctuations because 
of  the limited number of samples (photon pairs) used for reconstruction by quantum 
state tomography (QST)\cite{Jam2001}.

The experimental density matrices are obtained using the experimental setup depicted in 
Figure 1(a). We used a pair of BBO crystals pumped by a CW diode laser at 405 nm to generate 
the polarization entangled state $|\psi \rangle = (|H;H \rangle_{a,b} + |V;V \rangle_{a,b})/\sqrt{2}$. 
The generated photon pairs are  measured using 16 measurement setups, which are a 
combination of the bases for both of the photons\cite{Jam2001}. The measurement outcome  
$(n_k)_{i}$ ($i=1, \ldots, 16$), for which 1077 $\pm$ 33 photon pairs contribute, is converted 
to a density matrix $\hat{\rho}_k$ using the conventional method of QST, including maximum 
likelihood estimation. For the density matrices of the erroneous states, we experimentally 
obtained the measurement outcomes using the three input states 
$|\psi \rangle, |H;H\rangle, |V;V\rangle$ separately and added them together so that the off-diagonal 
terms of the density matrices are reduced by about 0.1 from that of the pure entangled state. 
We prepared 300 different density matrices for the normal state and 50 density matrices 
for the erroneous state. The average of the density matrices for the normal and erroneous 
states are shown in Figs. 1(b) and 1(c), respectively.

To compare the performance of  \pname \ in the experiments with that for their 
ideal cases containing some decohered states, we numerically simulated the 
measurement outcomes of the density matrices $(n_k)_{i}$ ($i=1, \ldots, 16$) with the 
contribution of 1000 photon pairs for both the `normal state' $|\psi \rangle$ and `erroneous 
state', where the off-diagonal elements of the density matrices are reduced by 0.1 similarly 
to the experiments. Then, the density matrices were calculated using the same QST method. 
In the computer simulation, we prepared 30,000 different density matrices for the normal 
state and 5,000 density matrices for the erroneous state.

\begin{figure*}
\centering
\subfigure[Raw elements and distances by the naive method]{
\includegraphics*[width=6.9cm]{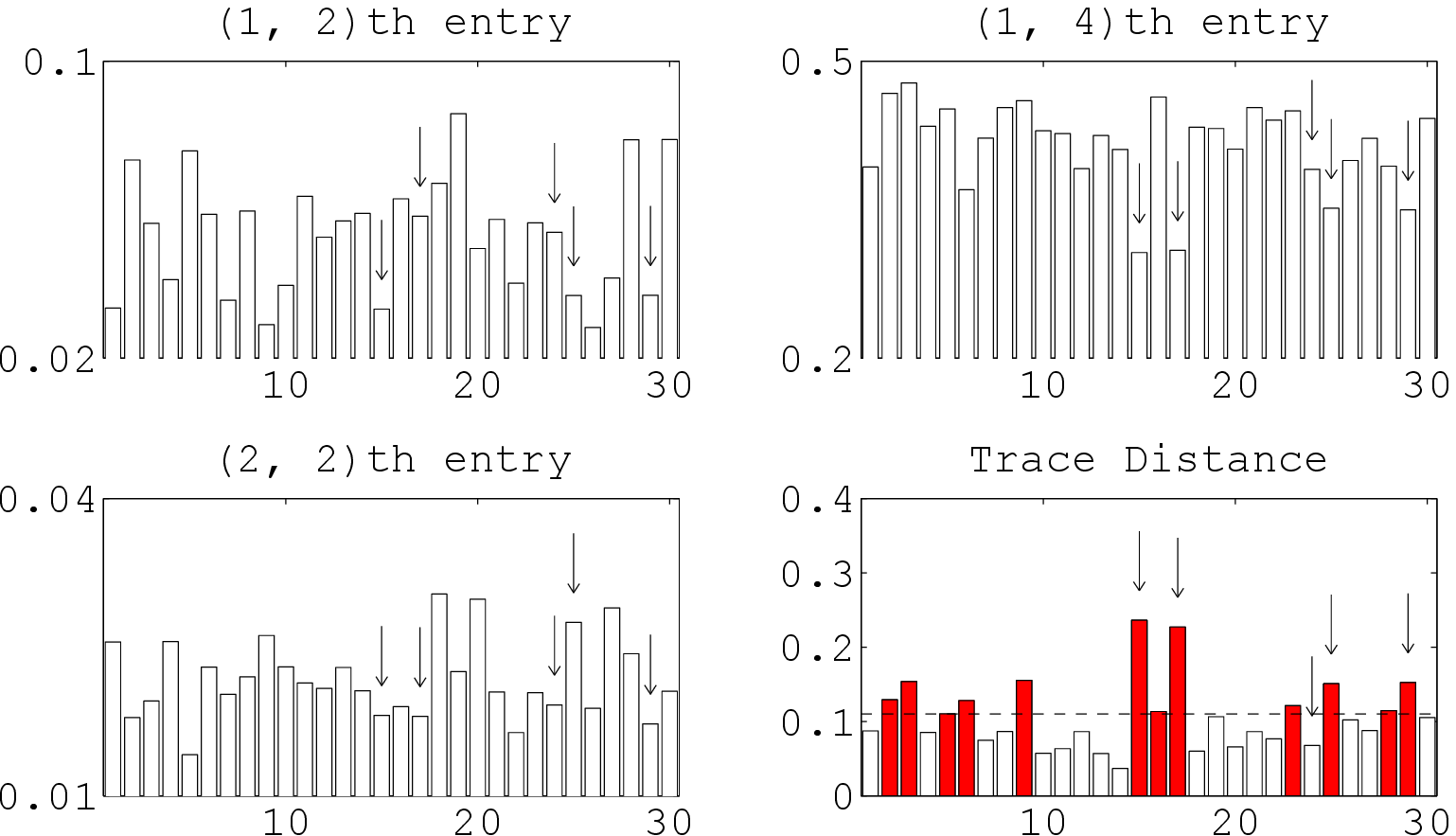}
\label{fig:hist_real_org}}
\subfigure[Estimated elements and distances by \pname]{
\includegraphics*[width=6.9cm]{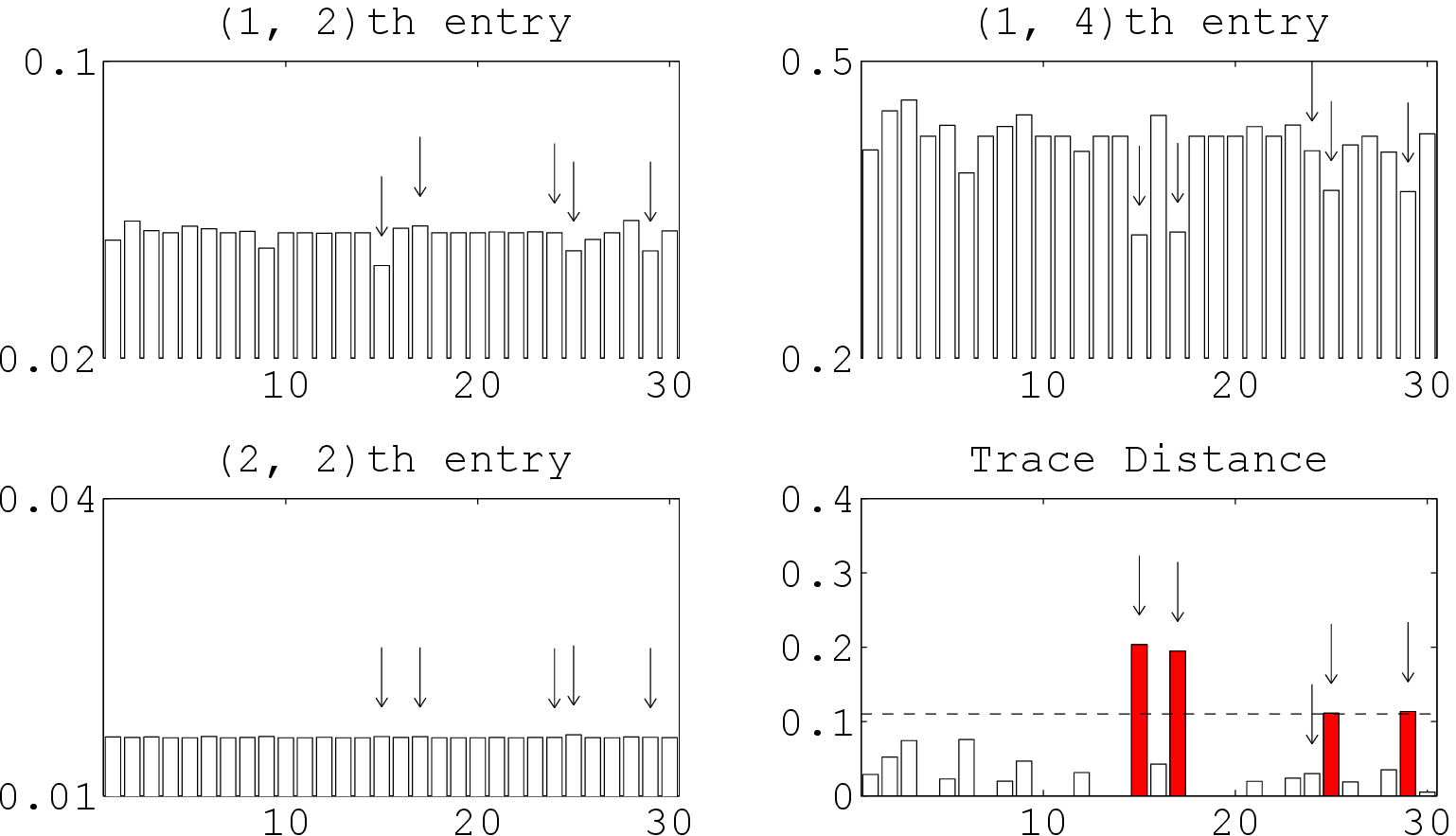}
\label{fig:hist_real_proposed}}
\subfigure[ROC curve]{
\includegraphics*[width=3.2cm]{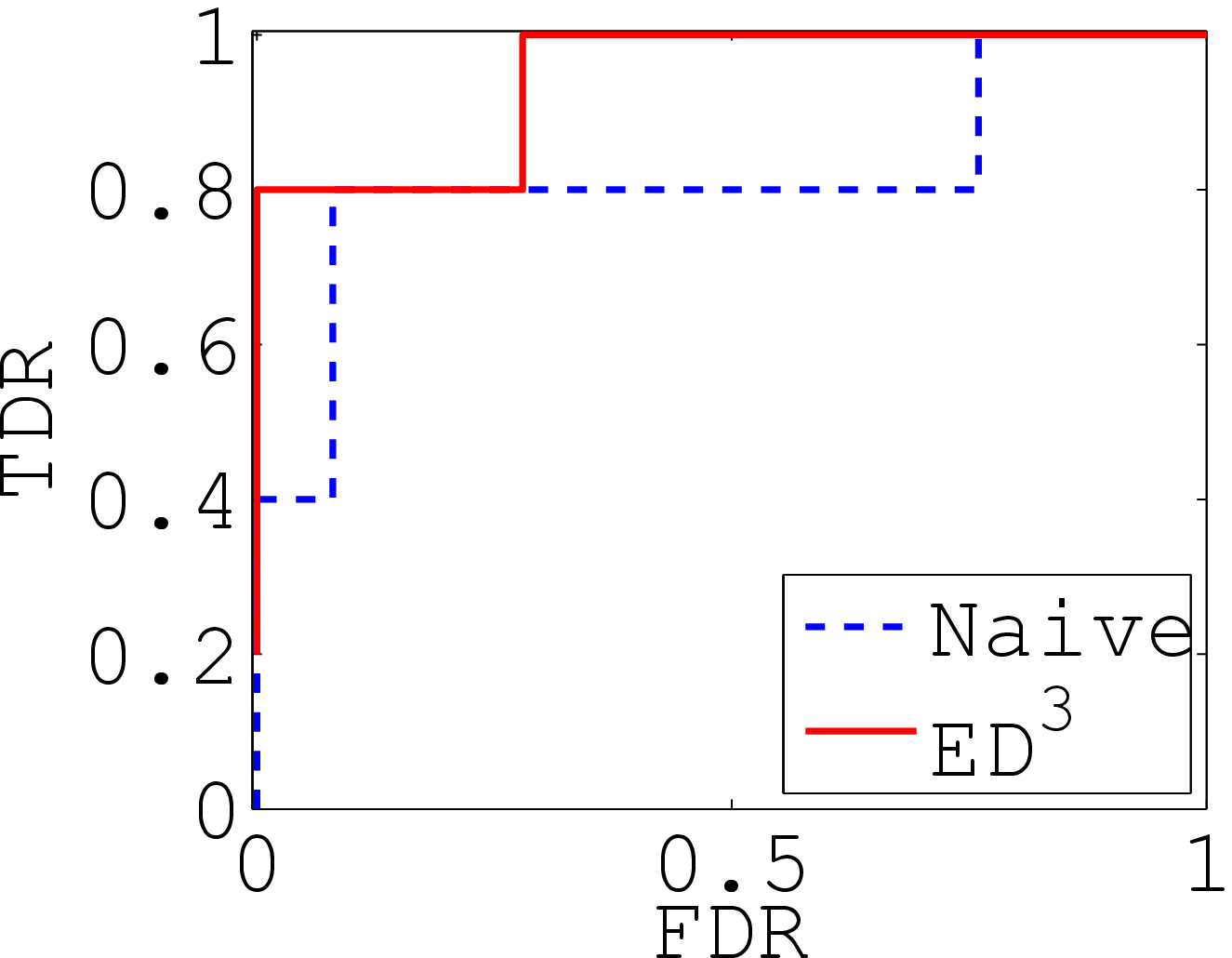}
\label{fig:roc_real}}
\caption{Performance analysis using a experimentally obtained dataset. (a) The 
raw values of the elements (1,2), (1,4) and (2,2) and  and the trace 
distance of the 30 density matrices in a test dataset. The arrows denote 
the artificially
introduced erroneous cases. (b) Corresponding elements and the trace 
distance of the estimated density matirices using proposed method (\pname).
The colored trace distances represent the cases where the state density 
matrices are judged to be erroneous under a threshold level, which is 
set to 0.11, depicted by the dashed lines in trace distance plots in
Figs. 4(a) and 4(b).
(c) The ROC curves for the naive method and \pname.}
\end{figure*}

\begin{figure}
\begin{center}
\includegraphics*[width=8.5cm]{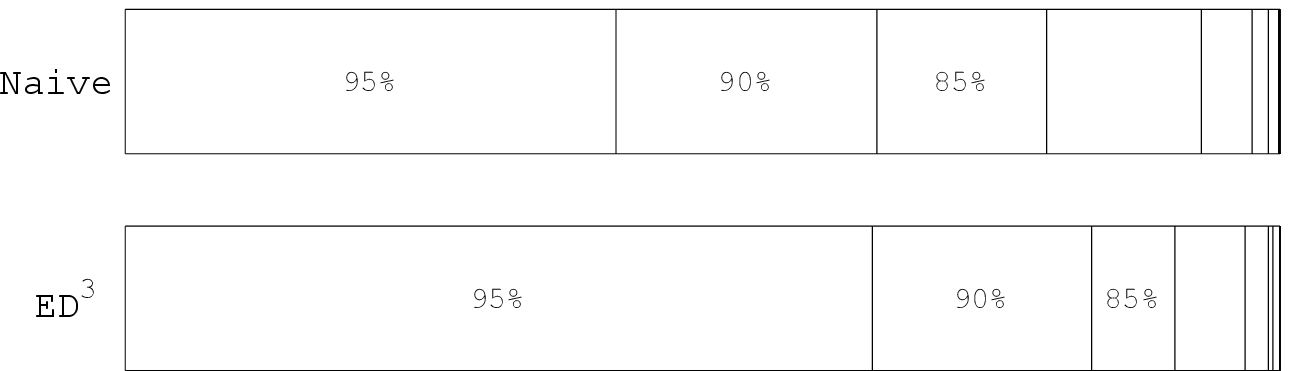}
\caption{Area under curve distribution over 1000 experimental datasets ({\it i.e.} randomly selected 25 matrices for
normal states and 5 matrices for erroneous states out of 300 and 50 experimentally 
obtained matrices respectively)}.
\end{center}
\label{fig:bar_auc_real}
\end{figure}


For performance evaluation, we obtained datasets each consisting of 25 
matrices randomly sampled from the normal states  and five matrices from the 
erroneous states for both the simulations and the experiments. For \pname, we 
prepared 10 preliminary datasets and tuned the parameter $\gamma$ so that the 
five erroneous matrices were discriminated well in every dataset. We then fixed 
the value of $\gamma$ and used it for the evaluations.

\section{Results}

An example of the evaluations using a computer-generated dataset is shown 
in Fig. 2. Figures~2(a) and 2(b) show the (1, 2)th, (1, 4)th, and (2, 2)th elements of raw density 
matrices $\hat{\rho}_k$ and density matrices $\theta+\omega_k$  ($k=1, \ldots, 30$) from the naive method and those
estimated by \pname \, respectively.
The arrows denote 
the artificially introduced erroneous cases.
Note that the trace distances are used for both naive method and ED$^3$ for a fair comparison.
These graphs indicate that \pname \ efficiently removes statistical fluctuations of the raw 
observed matrices and estimates the elements of the invariant $\theta$ for the normal 
matrices and the variated (1, 4)th elements of $\theta+\omega_k$ for the erroneous 
matrices. Figures~2(a) and 2(b) also show the trace distances Eq.(\ref{eq:td}) provided by the naive method and \pname, respectively. 
The colored trace distances represent the cases where the
state density matrices are judged to be erroneous under
a threshold level, which is set to $0.04$, 
depicted by the dashed lines in trace
distance plots in Figs. 2(a) and 2(b).
As easily 
understood by these figures, the true discovery rate (TDR) and the false discovery 
rate (FDR) of the erroneous matrices widely vary with the given threshold level where 
TDR and FDR are the rate of the true erroneous matrices (with the arrows) and the 
rate of the true normal matrices (without the arrows) in the matrices detected as 
erroneous (red colored), respectively. 
Figure~2(c) shows Receiver Operating Characteristic (ROC) curves obtained by changing 
the threshold level from 0 to the maximum trace distance for both methods, where 
the horizontal and vertical axes stand for FDR and TDR for every threshold level, 
respectively. Under this dataset, the curve of \pname \ is closer to the upper left corner 
(FDR $=0$ and TDR $=1$)  than that of the naive method,
indicating more reliable detection of the errors by \pname.

In order to obtain more quantitative comparisons between 
these curves for many different datasets, we use the area under curve (AUC) indicator, 
which is the percentage of the area under the ROC curve in the FDR--TDR plane and 
frequently used in data mining studies. It takes a value between 0\% and 100\% by definition, and a larger value 
indicates a better detectability of the error. Figure 3 shows the distribution of the AUC 
values over 1000 datasets and clearly demonstrates the superior performance of \pname \ 
in comparison with the naive method.

Finally, we test the performance of the proposed method using experimental datasets 
(Figs 4 and 5). Figures~4(a) and 4(b) respectively represent the (1, 2)th, (1, 4)th, 
and (2, 2)th elements of the raw density matrices and density matrices from the 
naive method and those estimated by \pname \, respectively.
The (1, 2)th and (2, 2)th raw elements in (a) contain some biases 
because the calibration accuracy of the phase shifters placed between single mode fiber 
and beam splitter, 
depicted in Fig. 1, is practically limited. These biases induced by the tolerated error of 
the experimental condition typically indicate that the approach based on the physical 
model of the `ideal' experiments does not account for the `true'  error detection, as 
mentioned earlier. On the other hand, these biases are properly reflected for the 
(1, 2)th and (2, 2)th elements estimated by \pname \ in (b). 
The graphs of the trace distances in Figs. 4(a) and 4(b) indicate better discrimination 
of the erroneous states by \pname \ than by the naive method, particularly in terms of 
FDR. The ROC curves depicted in Fig. 4(c) clearly indicate the higher detectability 
of \pname \ than the naive method for this dataset. Moreover, the distribution of 
the AUC values over 1000 experimental datasets ({\it i.e.} randomly selected 25 matrices for
normal states and 5 matrices for erroneous states out of 300 and 50 experimentally 
obtained matrices respectively) presented in Fig. 5 also demonstrates 
the better performance of \pname \ for the experimentally obtained data, similar to 
those generated by the computer simulation.

\section{Conclusion}

We have demonstrated that the proposed \pname \ based on a data mining method can more accurately detect small erroneous deviations in reconstructed density matrices that contain intrinsic fluctuations due to a limited number of samples, than a naive method checking the trace distance from the average of the given density matrices. A statistical analysis of the AUC over 1000 datasets of experimentally obtained and computer simulated density matrices clearly shows that ED$^3$ outperforms the naive method. We believe that quantum state data mining will be a key tool in broad area of physics where the detection of small deviations of quantum states reconstructed using limited number of samples are essential.

Note that here we are interested in the detection of anomalies of the quantum states itself. For instance, in this manuscript we are trying to detect the decohered states as anomalies. In this case, it is certainly important to perform the anomaly detection at the level of reconstructed density matrices, where the errors caused by the measurement apparatus have been compensated during the maximally-likelihood process. If one is interested in the errors caused by the measurement apparatus, the anomaly detection at a lower level may be more suitable. Note also that the anomaly detection at the level of density matrices are able to be applied to any quantum states represented by not only photonic qubits but also other many physical systems including superconducting circuits, trapped ions, and so on. In contrast, the anomaly detection at a lower level may be very dependent on the technical details of each physical systems (detection schemes, apparatuses, etc.) and thus not to be applicable as wide as ours. 
In this paper, we focused on the anomaly detection of decoherence in the quantum state.
The extension of \pname \ to the detection of arbitrary deviations including unintended unitary transformation is an interesting future study.

\section*{ACKNOWLEDGMENTS}

This work was supported in part by MEXT/JSPS KAKENHI Grant Number 21650029,
Quantum Cybernetics project of JSPS (No. 21102007), Grant-in-Aid from JSPS
(No. 23244079),
JST-CREST project, FIRST Program of JSPS, Special
Coordination Funds for Promoting Science and
Technology, Research Foundation for Opto-Science and
Technology, and the GCOE program.

\end{document}